\documentclass[a4paper,twocolumn,english,aps,prl,showpacs]{revtex4-1}
\usepackage[pdftex]{graphicx,color} 
\usepackage[T1]{fontenc}
\usepackage[latin1]{inputenc}
\usepackage{amsmath}
\usepackage{amsfonts}
\usepackage{color}

\makeatletter

\usepackage{babel}

\begin{document}

\title{Fractional Quantum Hall States in a Ge Quantum Well}

\author{ O. A.  Mironov$^{1,2}$,  N. d'Ambrumenil$^1$,  A. Dobbie$^1$, A. V. Suslov$^3$,
  E. Green$^4$ and D. R. Leadley$^1$}
\affiliation{$^1$Department of Physics, University of Warwick, Coventry CV4
7AL, UK \\
$^2$International Laboratory of High Magnetic Fields and Low
Temperatures, Wroclaw, Poland \\
$^3$National High Magnetic Field Laboratory, 1800 East Paul
Dirac Drive, Tallahassee, FL 32310,
USA \\
$^4$Hochfeld-Magnetlabor Dresden (HLD-EMFL), Helmholtz-Zentrum Dresden-Rossendorf, D-01314 Dresden, Germany}

\date{\today}

\begin{abstract}
Measurements of the Hall and dissipative conductivity of a strained Ge quantum well on a SiGe/(001)Si substrate in the quantum Hall regime are reported. We analyse the results in terms of thermally activated quantum tunneling of carriers from one internal edge state to another across saddle points in the long range impurity potential. This shows that the gaps for different filling fractions closely follow the dependence predicted by theory. We also find that the estimates of the separation of the edge states at the saddle are in line with the expectations of an electrostatic model in the lowest spin-polarised Landau level (LL), but not in the spin-reversed LL where the density of quasiparticle states is not high enough to accommodate the carriers required.
\end{abstract}
\pacs{73.20.Mf, 73.21.-b, 73.40.Hm, 73.43.Cd, 73.43.Lp}
\maketitle

The strongest fractional quantum states have been
found in delta-doped GaAs quantum wells and heterostructures.
These have been the cleanest samples with
the highest mobilities and yet the strength of a fractional
quantum Hall state does not appear to be connected in a
simple way with the zero magnetic field mobility of the samples involved \cite{Deng2014,Umansky09}.


The strength of a quantum Hall state is associated with how 
rapidly the dissipative conductance drops with
temperature.
This is known to be affected by 
the long-range
impurity potential, arising out of the ionised donors in a
semiconductor heterostructure, although
describing this effect quantitatively has proved difficult. 
The background potential leads to the formation of compressible
regions, which then partially screen the background
potential. Predicted theoretically in \cite{Efros88_2,Pikus-Efros94},
and later verified in
scanning electron transistor measurements \cite{Martinetal05}, these
compressible regions are
separated from the percolating incompressible region by internal
edges. The transfer between these internal edges is then the principal
dissipative process at low temperatures \cite{Polyakov94,dABIHMorf11}.  Characterising these
internal edges is not just an important challenge for describing 
the quantum Hall response, but
also to the interpretation of Aharonov-Bohm
interference experiments at filling factor $\nu=5/2$ \cite{Willett2013}. It turns
out, for example, that it is crucial to the interpretation of these experiments to know 
whether any charge redistribution associated with changing the field
occurs internally (by shifts in the local occupation of compressible
regions) or by the outer sample edges \cite{Keyserlingk15}.

Here, we report conductivity measurements on a p-type strained germanium
quantum well. 
There is a clear family of quantum Hall states in the composite
fermion (CF) series centered on $\nu=1/2$ \cite{Jain89}.  Our analysis of the data shows that the model can give a consistent fit
across all filling fractions in the CF family for the gaps in these
systems and explains why the zero field mobility is not simply
connected with the strength of the quantized Hall states.  
The gaps are in line with the predicted dependence on filling fraction. In addition we estimate the typical separation at a
saddle point, $a$, of two compressible regions occupied by the same type of
carrier. As a function of the gap, these correlate
well with the separation of neighboring compressible regions
containing carriers of opposite charge, which can be predicted
on the basis of an electrostatic model
\cite{Chklovskii92}. While the dependence of the width parameter, $a$,
on the gap, $\Delta_s$,
is the same for filling fractions in the lowest
Landau level (LL) above and below $\nu=1/2$, there is a systematic
shift to higher values for the quasihole states ($\nu >
1/2$). However, for states above $\nu>1$ where multiple LLs are occupied, the dependence does not
coincide with the predictions of the electrostatic model. We
attribute this to the need for a charge rearrangement across the
incompressible region, which is too large for a simple edge to accommodate.

Our sample was grown by reduced pressure chemical vapour deposition at
the University of Warwick (sample ID 11-289SQ1D). It is taken 
from the same wafer that was reported in \cite{Dobbie12}, it
has a density of $2.9 \times 10^{11}$ cm$^{-2}$ and a mobility of $1.3
\times 10^6$ cm$^2$/Vs (with current flow along the $[110]$ direction). 
The hole effective mass of 0.073(1)$m_e$, measured from the
temperature dependence of the Shubnikov-de Haas oscillations at low
field, is remarkably similar to that of electrons in GaAs ($0.067
m_e$) and much lower than for holes in GaAs. The Dingle ratio is $78\pm 2$ \cite{Mironov14,Supp}.

\begin{figure}[t]
\includegraphics[width=3.2in]{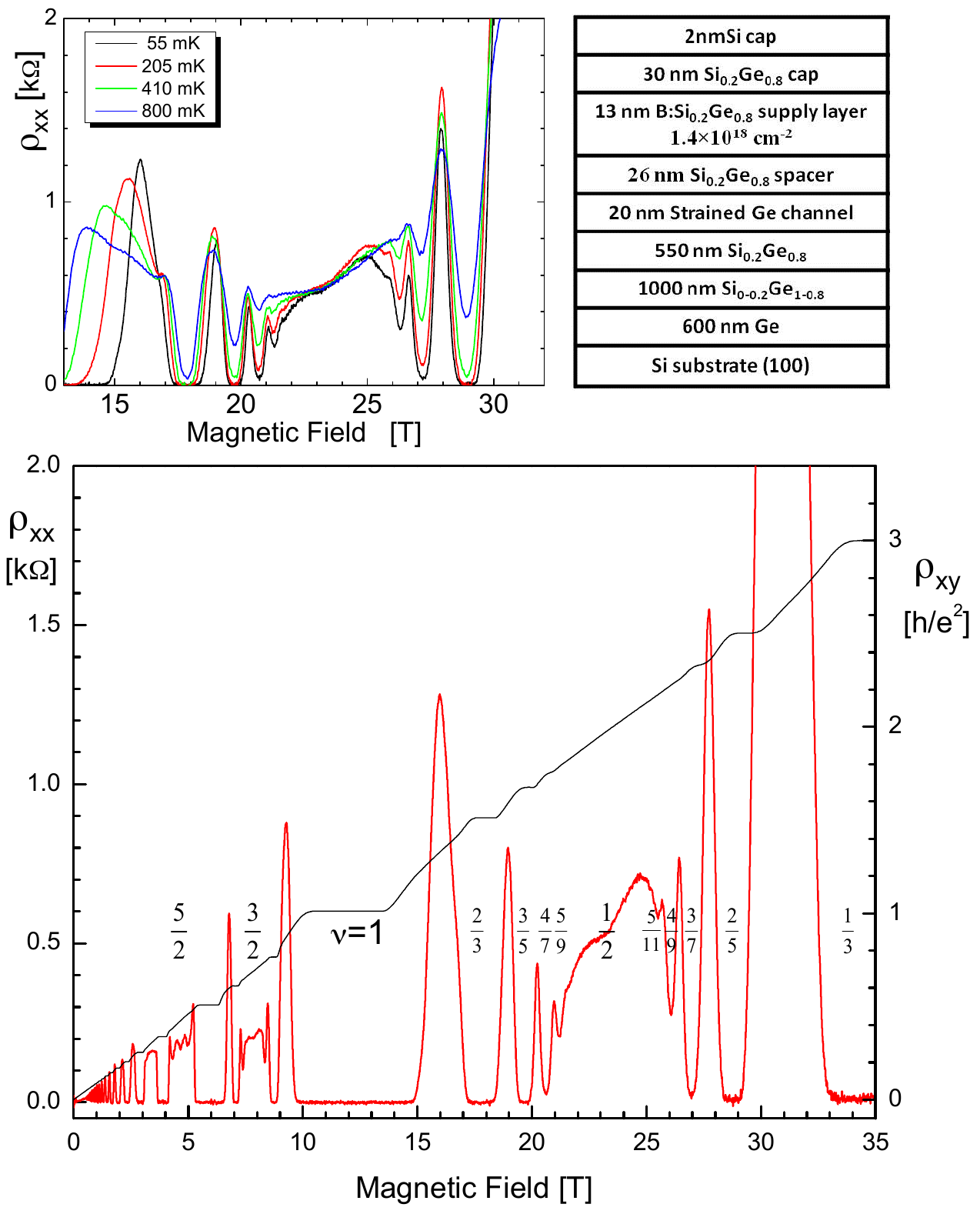}
\caption{\label{fig:Resistance} 
The Hall resistance, $\rho_{xy}$, and dissipative response,
$\rho_{xx}$, at 55mK (main figure) and the dissipative response at different
temperatures \cite{Supp}
in the high field regime (top left). The wafer structure is
summarized in the schematic (top right).} 
\end{figure}

Fig. \ref{fig:Resistance} shows the structure of the wafer. 
A reverse linear graded, strain tuning buffer \cite{Shah08}
terminating in over-relaxed Si$_{0.2}$Ge$_{0.8}$, is followed by
the Ge
quantum well which is under 0.65\% biaxial compressive strain.
Holes are supplied from a boron doped layer that is set back 26 nm
above the quantum well. Magnetoresistance measurements were performed
at the National High Magnetic Field Laboratory  using static fields of up to 35 T and at temperatures down to
26 mK.
We note that results on a similar sample grown
in our laboratory have already been reported
\cite{ShiMyronovCF15} but without any discussion of the CF family in the lowest
Landau level.

Figure \ref{fig:Resistance} shows field-traces of the longitudinal
resistance taken in a $4\times4$ mm square geometry at four different
temperatures with the current along the $[110]$ direction. The Hall
resistance was measured separately in a Hall bar geometry. 
There are clear quantum Hall states at
fractions in the main CF sequence around $\nu=1/2$ and
$\nu=3/2$. As seen in CdTe \cite{Piot10}, there is some deviation 
from simple activated behavior even when there are good quantum Hall states. 
Around $\nu=5/2$ we find a minimum on $\rho_{xx}$ but no
clear plateau at $\nu =5/2$. The mobility of
our sample is high for a Ge sample, but it is still significantly less than for
GaAs samples showing such clear FQH states. For example in a p-doped GaAs sample,
with $\mu = 2.3\times 10^6$ cm$^2$/Vs (close to twice that of our sample),
there are precursor signals of states at $\nu=3/5$ and at $\nu=4/7$ but not 
quantized Hall states \cite{WatsonCsathy12}. In n-doped GaAs no
sign of the $\nu=5/2$ state is visible for  $\mu \lesssim 6.7 \times
10^6$ cm$^2$/Vs \cite{Deng2014,Umansky09}.

While the zero field mobility
reflects all scattering, we assume that it is the long-range
potential of the ionized donors that controls the dissipation in quantum
Hall states. Any short range scattering centres, not located directly in the saddle points of
the long range impurity potential, will not affect the response of
the quantum Hall state. This makes clear why mobility is unlikely to
be a good indicator of strong quantum Hall states \cite{Deng2014}.

\begin{figure}[t]
\includegraphics[width=3.3in]{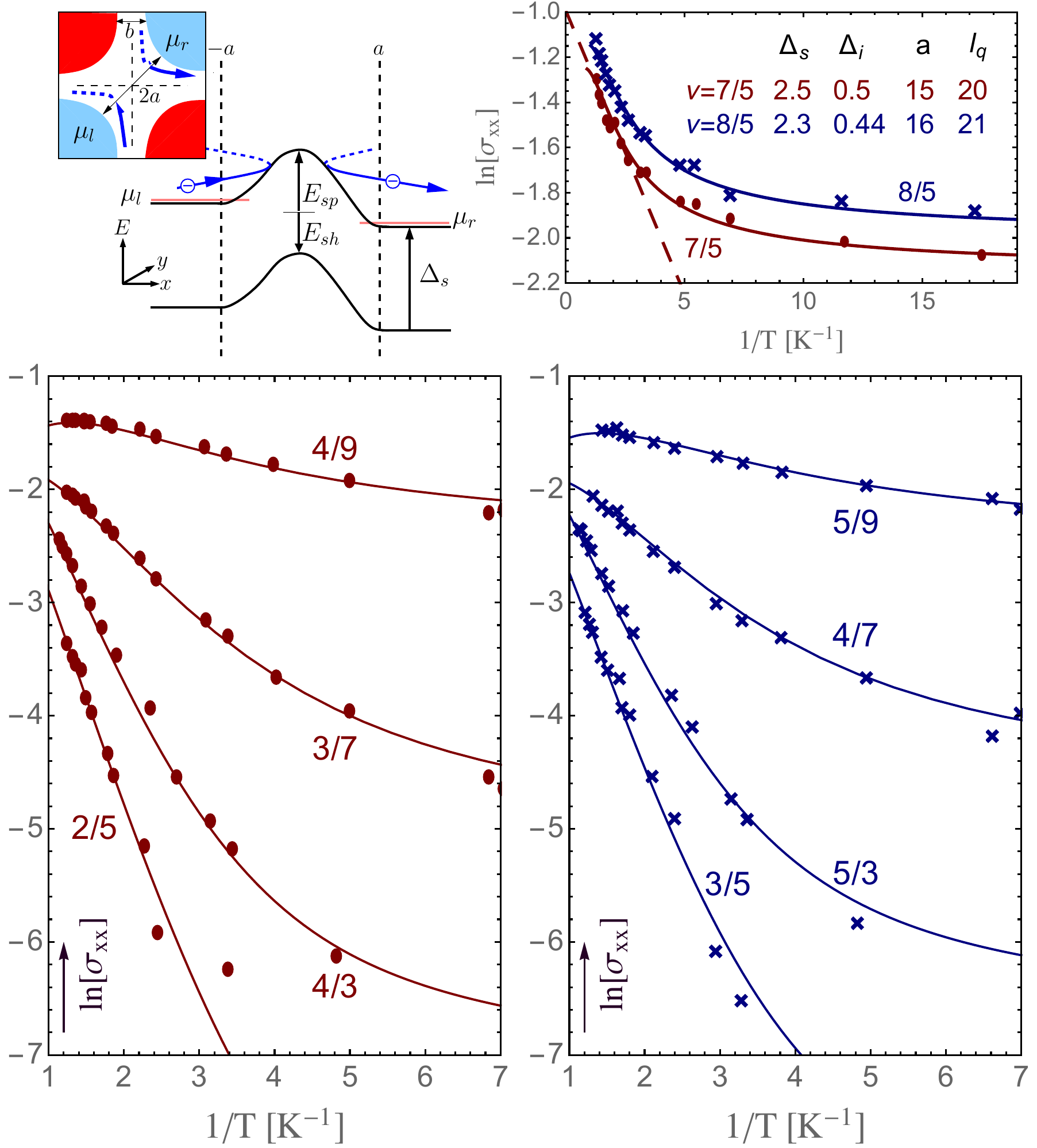}
\caption{\label{fig:Model} 
Band alignment and particle flow across a typical saddle point in the
potential (top left).  The blue/red 
regions are compressible with nucleated QPs/QHs, while the white
region is incompressible.  
Dissipation occurs when excitations move across a saddle point from
a compressible region with chemical potential,
$\mu_l$, to compressible region at $\mu_r$.  The transfer proceeds
by thermally activated tunneling across the energy barrier, which on
average is $\Delta_s/2$. 
The resulting logarithmic conductance  (with $\sigma$ in units of $2(qe^2)/h$) at different filling fractions
is shown top right and below. For the $\nu=7/5$ and $\nu=8/5$ states,
the theoretical prediction and model parameters used to
describe the data  are shown (the data at $\nu=8/5$ have all been raised by 0.1
for clarity). }
\end{figure}

We have analysed the data using the model developed in
\cite{dABIHMorf11,dAMorf13}. This assumes that 
localized regions or puddles of compressible regions are nucleated
within the incompressible quantum Hall fluid. 
The dissipative
response 
is controlled by excitations, localized in one puddle, crossing
to another via a saddle point in the impurity potential, see 
Fig.~\ref{fig:Model}. The saddle points act as effective resistors and
the puddles of quasiparticles (QP) and quasiholes (QH) act as reservoirs
in a resistor network~\cite{Polyakov94}. The response is then 
that of the average saddle point (with barrier height $\Delta_s/2$) \cite{Dykhne71,Asman15}.

If the energy required to create a QP/QH pair near a
saddle is $\Delta_s$, the average barrier height to traverse a saddle
is  $\Delta_s/2$ for both QPs ($E_{sp}$) and QHs ($E_{sh}$). In the absence of
tunneling through the saddle point, the dissipative conductance per
square is $ [2(qe)^2/h]\,e^{-\Delta_s/2kT}$ as predicted by Polyakov
and Shklovskii  \cite{Polyakov94}. Taking
account of tunneling gives a dissipative conductance per
square 
\begin{equation}
\sigma_{xx}^\Box = 2\frac{(qe)^2}{h} F[\Delta_s, a/l_q],
\label{eq:Conductance}
\end{equation}
where $a$ is the typical saddle point width (see Fig.~\ref{fig:Model})
and $l_q$ is the magnetic length for QPs/QHs with fractional charge
$qe$ \cite{dABIHMorf11,Fertig_Halperin87,Supp} . We fit the data to the computed form
for $F$ 
using
the gap, $\Delta_s$, and the width parameter, $a$, as free
parameters.

We convert the measured minimum values of
$\rho_{xx}$ at each filling fraction and for each temperature 
to equivalent conductivities $\sigma_{xx}$ (the Hall resistance is
taken as the corresponding quantized value). 
The 
results for $\nu=7/5$ are shown in Figure \ref{fig:Model}. We also
show the results of the more traditional method of assuming the
Arrhenius form, drawing a tangent to the curves at the inflection
point and identifying the gradient with an energy gap, $\Delta_i$. 

The aspect ratio (width/length) of
the active region of the sample gives an overall additive constant to $\ln
\sigma_{xx}$ \cite{Supp}. We use this as a consistency check on the model as it should not vary
significantly between filling fractions.
For the states at $\nu=2/3, 2/5, 3/5, 3/7$ and $4/7$, the data in Fig.
\ref{fig:Model} fitted
assuming a constant aspect ratio of 1.8. At filling fractions
with small gaps, we can expect that the effective width of the
percolating incompressible region may reduce as the state is
weaker. At $\nu = 4/9$ and $\nu=5/9$
the aspect ratio is reduced to 1.4 and 1.5 respectively. However,
we should emphasize that at these filling fractions the states are
weak (the ratio between low and high temperature conductance is less
than 1.9).  


\begin{figure}[t]
\includegraphics[width=3.2in]{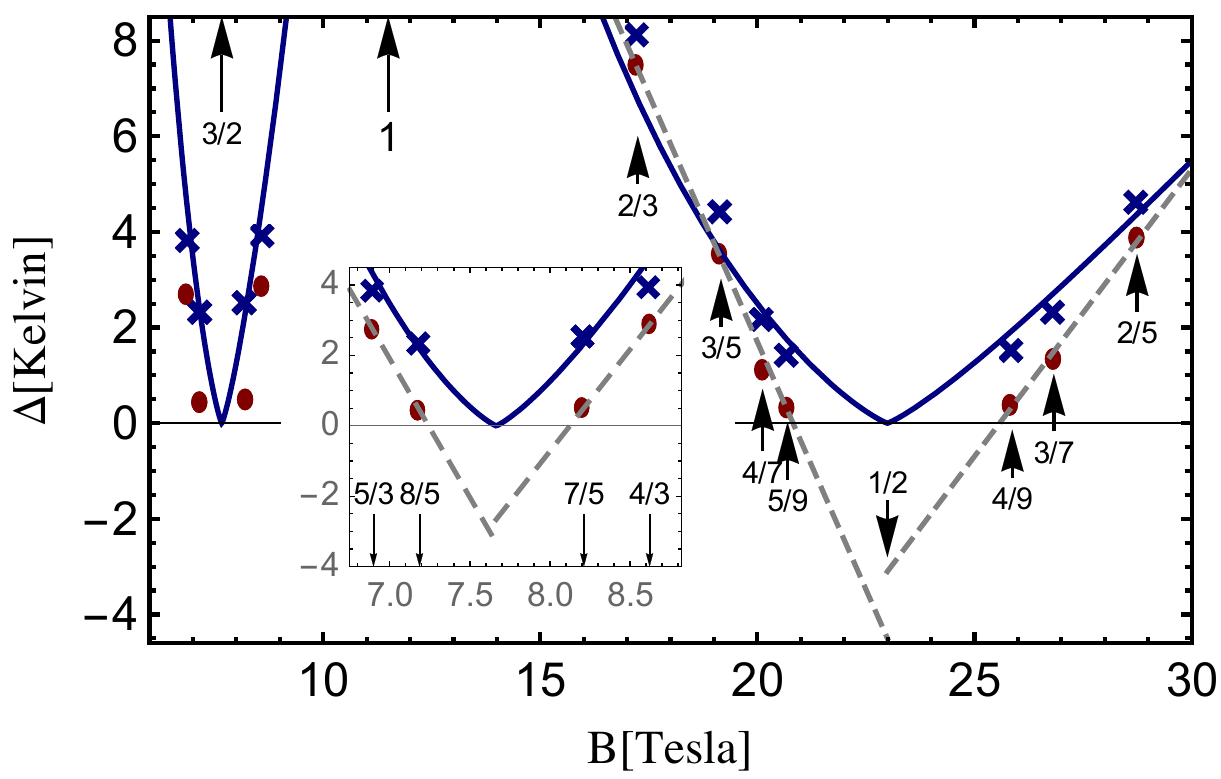}
\caption{\label{fig:GapAnalysis} 
Estimates of the gap as a function of magnetic
field. 
The crosses denote the estimate of the gap
at each $\nu$, 
$\Delta_s$. The ovals show the energies, $\Delta_i$, estimated from the 
gradients in the Arrhenius plots
taken at the point of inflection for each $\nu$ (see
Fig.~\ref{fig:Model}) . 
The solid lines are the dependence predicted in CF theory \cite{HLR}
with $C'=2$ and $C=0.25$, see (\ref{eq:CFTheory}). The dashed straight lines are 
 best linear fits. Insert: Results close to $\nu=3/2$
on an expanded scale. }
\end{figure}

In Fig.~\ref{fig:GapAnalysis} we compare the results for $\Delta_s$
with the predicted dependence on magnetic field for the composite fermion
model \cite{HLR}. For spin-polarised states at filling fraction
$\nu_p=p/(2p+1)$, $1\pm \nu_p$ or $2-\nu_p$, the theory predicts that
the gap in a homogeneous system, $\Delta_h$, varies as 
\begin{equation}
\Delta_h = \frac{C}{|2p+1|(\ln|2p+1| + C')}
\frac{e^2}{\epsilon l_0},
\label{eq:CFTheory}
\end{equation}
where $C$ and $C'$ are
dimensionless constants and $l_0$ is the magnetic length. Analysis of the logarithmic divergences as
$p\rightarrow \infty$ for a homogeneous system with zero width and
without LL-mixing corrections suggested $C=1.27$ \cite{HLR} while
comparison with exact diagonalization studies put $C'=3.0$ and, later,
$C'=4.11$ \cite{Morf_NdA_SdS_02}.  We have treated $C$ and $C'$ as free parameters and compared
with the values we obtain for $\Delta_s$. The results are shown
as solid lines in Fig.~\ref{fig:GapAnalysis} with $C=0.25$ and $C'=2$. 

The agreement between CF theory and the estimated gaps, $\Delta_s$, is good
even at filling fractions above $\nu=1$.  The values of the constants $C$ and $C'$ are
different from those estimated for the homogeneous zero width case and
indicate a larger role for the logarithmic corrections to the CF
effective mass.
The result
(\ref{eq:CFTheory}) can be rewritten as a formula for the effective
mass via 
\begin{equation}
m^*(p) = \hbar^2 \left(\frac{\epsilon l_0}{e^2}\right) \frac{\ln|2p+1| + C'}{C}.
\label{eq:effMass}
\end{equation}
As expected the effective
mass depends on the magnetic length---it varies between 0.69 $m_e$ at $\nu = 5/3$ to 1.09 $m_e$
at $\nu = 2/3$ (both have $p=-2$) and on $p$. Our results also suggest
that the dependence of the effective mass on $p$ is stronger than predicted
for the homogeneous 2DEG ($C'=2$ instead of 4.1 suggested by diagonalizations
of the Hamiltonian of homogeneous systems
\cite{Morf_NdA_SdS_02}).

Although there is no microscopic model giving the typical saddle point
width, $a$, as a function of
$\Delta_s$, we assume that it should be related to $b$, the width of
the incompressible region between two puddles (see upper left pane in
Fig. \ref{fig:Model}).
We estimate $b$ using the theory developed for edges, which
computes the electrostatic potential induced by fixing the charge density at
the incompressible value \cite{Chklovskii92}.  
The width, $b$, is determined by setting the energy to create a quasiparticle-quasihole (QP-QH) pair equal to
the electrostatic energy gained by transferring a QP from the low to
the high density side of the incompressible region. This gives
$b^2 =  \frac{8  e(e/q) \Delta}{q\,\pi\, dn/dx|_c}$
where $dn/dx|_c$ is the gradient of the carrier density at the
centre of the incompressible strip if
the system were fully screening of the background
impurity potential.

\begin{figure}[t]
\includegraphics[width=3.2in]{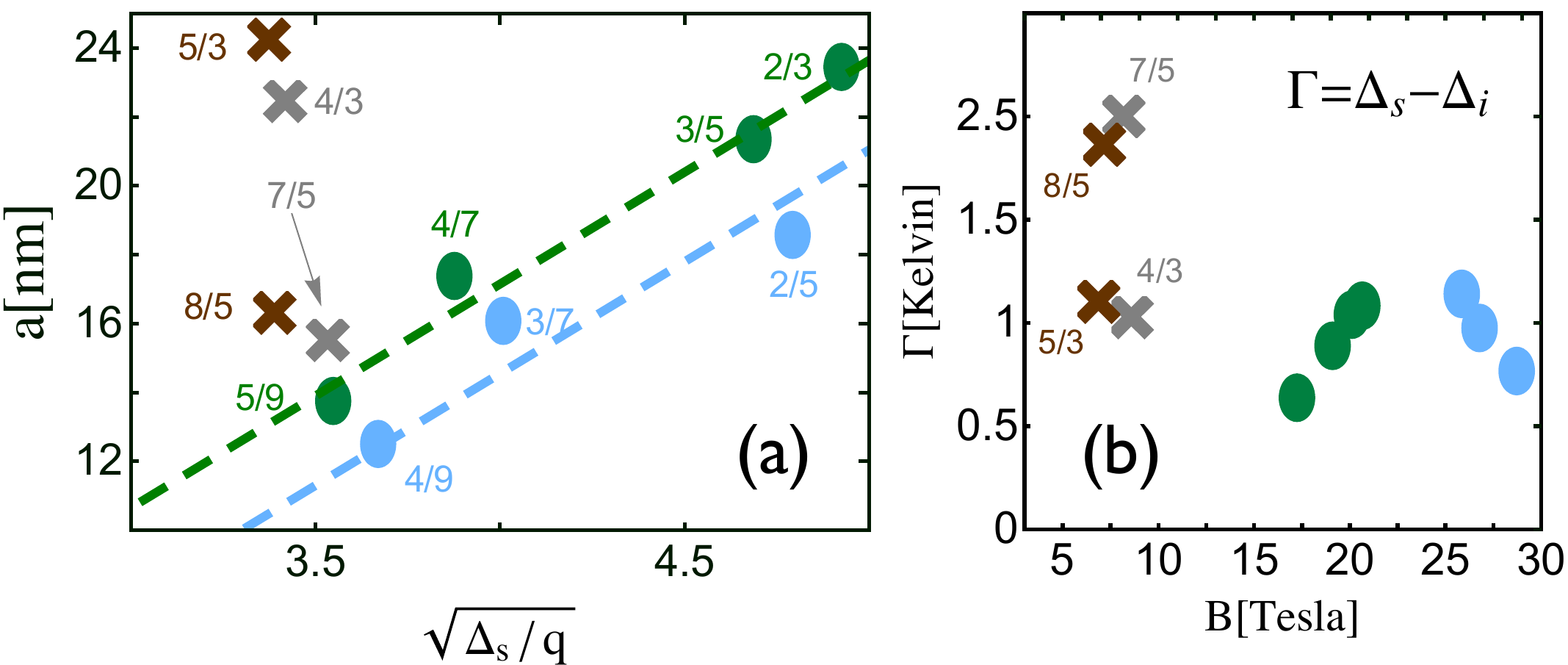}
\caption{\label{fig:AandGamma} (a) The saddle point width parameter, $a$,
plotted against $\sqrt{\Delta_s/q}$ (in units of K$^{1/2}$), for the
different CF families (left). Dashed lines are
  guides to the eye. (b) $\Gamma = \Delta_s - \Delta_i$ for
  different filling fractions. $\Gamma$ reflects the reduction in the apparent gap due to
tunneling and has traditionally been assumed independent of filling factor.
}
\end{figure}
In Fig.~\ref{fig:AandGamma}(a) we show $a$ as a function of
$\sqrt{\Delta_s/q}$ where the $\Delta_s$ are values from fits to
the data. For the two families $\nu^{qp}=p/(2p+1)$ and $\nu^{qh} =
1-p/(2p+1)$, we see that a linear dependence on $\sqrt{\Delta_s/q}$
works quite well. There is evidence of a systematic difference between
the QP family and the QH family, with the width in the QH case slightly larger than in the QP case. 
A difference between the two cases is not surprising, given that the
boundary of a compressible puddle is an internal edge and the
fractionally charged QP/QH is an internal edge excitation. 
As the model
does not factor in any details of the edge state reconstruction,  a small systematic difference between the
results for the two families ($\nu= p/(2p+1$ and $\nu = 1-p/(2p+1)$)
is to be expected.

In the case of the partially occupied reverse spin LL (RSLL), 
Fig.~\ref{fig:AandGamma}(a) shows
that $a$ does not follow the simple CSG scaling \cite{Chklovskii92}. The
scaling assumes that the compressible regions behave like perfect
metals, {\it i.e.\/} that the screening length in the puddles is short enough for
the electrostatic model to be a good approximation. This turns out to be unlikely
in the RSLL in our sample. The theory would require that the screening electron density changes between
edges of an incompressible
strip by $\Delta n \sim b\, dn/dx|_c$.  At filling fraction, $\nu =
1+p/(p+1)$, the
corresponding change in
number density of QPs, with charge $qe=e/(2p+1)$, is 
$\Delta n_{QP} = \Delta n / q$. The average density of electrons in
the RSLL is $n_{1} = n\, p/(3p+1)$. 
This gives:
\begin{equation}
\frac{\Delta n_{QP}}{n_{1}} = \frac{\Delta n}{n}  \frac{(2p+1)
  (3p+1)}{p}.
\label{eq:FracofQPs}
\end{equation} 
For uncorrelated ionised donors, the typical density
gradient would be 
$dn/dx|_c \sim \sqrt{ n/8\pi } /d^2 $ \cite{Pikus-Efros94}, which would give
$\frac{\Delta n_{QP}}{n_{1}} \sim 0.2$ at $\nu=4/3$ (p=1) and 0.3 at 
$8/5$ (p=2). These would be large density variations in a LL and would take
the system close to (or into) 
neighboring CF states (in the CF model, a ratio 1/(p+1) defines the next CF state in
the hierarchy). The assumption of fully
screening metallic regions, which is implicit in the CSG model,
applies to this sample at filling fractions above $\nu=1$.

Our model is not consistent with
an approach, which assumes a $\nu-$independent broadening due to
impurities. The difference, $\Gamma=\Delta_s - \Delta_i$, between the intrinsic gap and
the slope measured in Arrhenius plots of the dissipative conductance,
depends on the strength of the incompressible state. Fig
~\ref{fig:AandGamma}(b) shows the broadening parameter,
$\Gamma$,
which, according to our model, varies between 2.5K at $\nu = 7/5$ and around 0.6K at $\nu =
2/3$. $\Gamma$ is smaller the larger the intrinsic gap. 
On the other hand, the dashed lines in Fig.~\ref{fig:GapAnalysis}(a) show that,   if we fit 
$\Delta_i$ by a linear
function of $B-B(\nu=1/2)$ \cite{Jain89}, 
we obtain different values for the broadening
and effective masses: 
4.5K and 0.65$m_e$ for $\nu<1/2$, and 3.1K and 1.1$m_e$ for $\nu>1/2$.
We conclude that estimates of intrinsic gaps using states of significantly
different strengths on the basis of a $\nu-$independent
broadening parameter are likely to be unreliable, see also \cite{Samkharadze11}.

We have reported measurements of the dissipative response and the Hall
response of a Ge quantum well
in the fractional quantum Hall regime. The results  across all filling
fractions, for which FQH states are found, fit well with the predictions
of a model of thermally assisted quantum tunneling. The model demonstrates that
the properties of the internal edge states around puddles
control how a fractional quantum Hall state responds to
any slowly varying background potential.
Our sample has a high mobility
for Ge but still significantly below that for GaAs samples (both p- and
n-type) but shows much stronger quantum Hall states than would be
expected given the mobility. Our
model makes clear why the mobility is not likely to be a
reliable measure of the sample in the quantum Hall regime. The strength of the quantum
Hall state is determined by the strength of the long-range impurity
potential at saddle points in the potential.

We thank R. H. Morf, R. Morris and M. Myronov for helpful discussions. 
We acknowledge support from EPSRC grants EP/J001074 and
EP/F031408/1. The National High Magnetic Field Laboratory
is supported by 
NSF Cooperative Agreement No.
DMR-1157490 and by the State of Florida. We thank 
G. Jones, S. Hannahs, T. Murphy, J.-H. Park, and M. Zudov
for technical assistance with experiments at NHMFL. 
OAM thanks M.Uhlarz, M. Helm and J. Wosnitza for the support of the 
initial HLD-HZDR 25 mK/16T measurements in 2012 and the National Scholarship 
Program of the Slovak Republic for the Researchers Mobility Support in 2013.

\bibliographystyle{apsrev4-1}
\bibliography{SiGeFQHE}
\end{document}


\title{Fractional Quantum Hall States in a Ge Quantum
  Well---Supplementary Information}

\author{ O. A.  Mironov$^{1,2}$,  N. d'Ambrumenil$^1$,  A. Dobbie$^1$, A. V. Suslov$^3$,
  E. Green$^4$ and D. R. Leadley$^1$}
\affiliation{$^1$Department of Physics, University of Warwick, Coventry CV4
7AL, UK \\
$^2$International Laboratory of High Magnetic Fields and Low
Temperatures, Wroclaw, Poland \\
$^3$National High Magnetic Field Laboratory, 1800 East Paul
Dirac Drive, Tallahassee, FL 32310,
USA \\
$^4$Hochfeld-Magnetlabor Dresden (HLD-EMFL), Helmholtz-Zentrum Dresden-Rossendorf, D-01314 Dresden, Germany}
\maketitle
\date{\today}

\maketitle

In resistivity measurements in the quantum Hall regime, it is
crucial to be able to have confidence in the measurement of the
temperature.  We used a RuO temperature sensor, which was mounted on
the probe near the sample. Both sample and thermometer were loaded
directly into the mixing chamber and thus were immersed in the
mixture, in order to secure equal temperatures for the
sample and thermometer. 

\begin{figure}[b]
\includegraphics[width=2.7in]{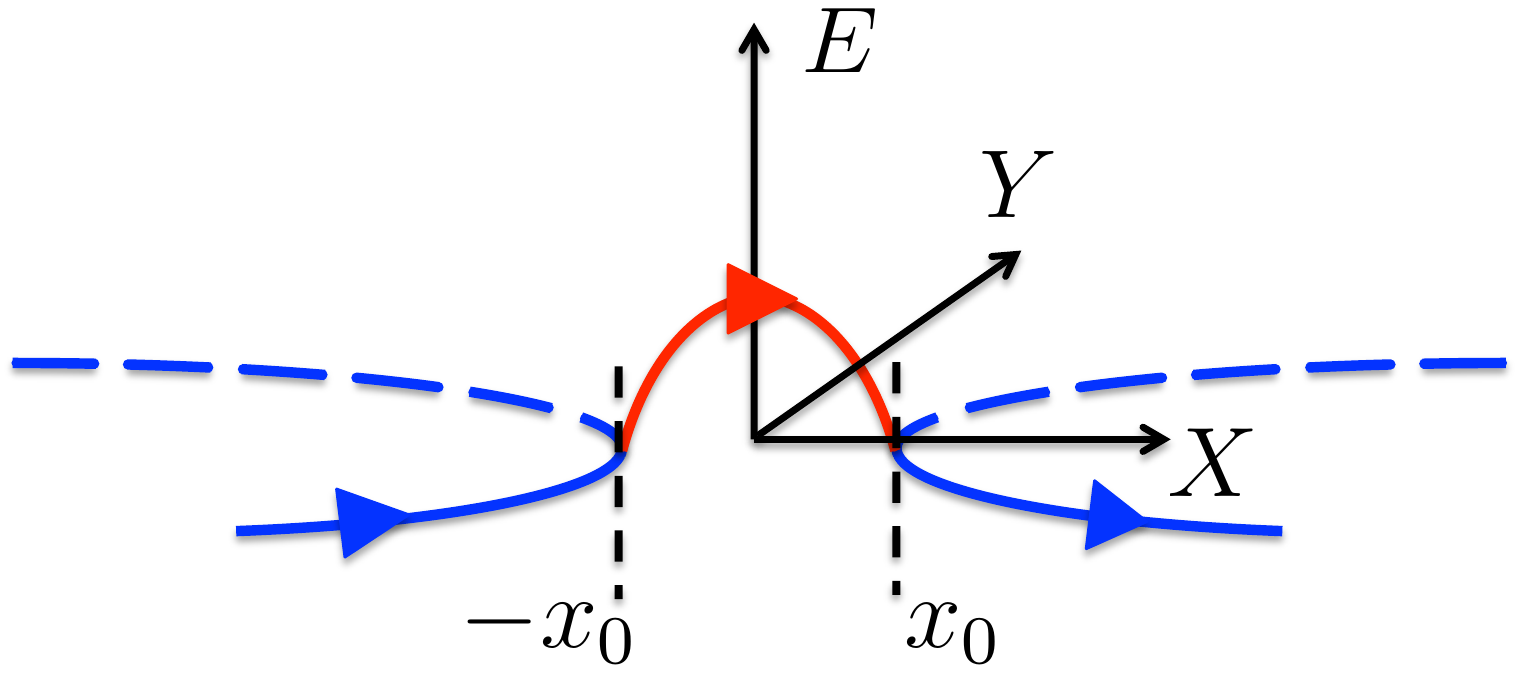}
\caption{\label{fig:Tunnel} 
The path of the cyclotron orbit center, ${\bf R} = (X,Y)$, of a QP as
it crosses a
saddle point. The QP drifts along the classical trajectory, which
encircles a local minimum of the impurity potential (left blue
line). At the saddle point, it can either continue in the orbit (blue
dashed line)  or
tunnel over the barrier and into the classical trajectory on the other
side. The tunneling trajectory (red line) can be modelled
within the WKB approach. It runs from $-x_0$ to $x_0$ the two points of closest
approach. The effective Hamiltonian is that of an inverted
parabola in one dimension, with momentum $Y$ and coordinate $X$.
The components of the cyclotron orbit coordinates, $X$ and $Y$, are conjugate variables with $[X,Y]=i l_q^2$
\cite{Fertig_Halperin87}, where $l_q$ is the QP magnetic length, and
hence $Y$ acts as the momentum to the position $X$.
}
\end{figure}

To avoid sample self-heating low current measurements of 20 nA and 50 nA
were used. As only the zero field calibration curve for this
thermometer is available and the magnetic field effect on this curve
is unknown, temperature stabilization was performed at zero magnetic
field by applying a constant power to the cryostat heater. The
magnetic field was swept up and back down to zero, a check made that
the temperature had not
drifted during the field sweep, and the temperature
then restabilized at another value. To avoid Joule heating and to 
keep the temperature deviation
in the field small, a plastic sample-thermometer holder and
small sweep rates for the magnetic field (1 T/min and
-2T/min) were used.
These procedures are one of the standard approaches used at the NHMFL 
for correct temperature measurements in high magnetic fields.

We have analysed the resistivity data on the basis that the long range
impurity potential, set up by the ionized donors, dominates the response in the quantum Hall
regime. An indicator that this is the case, is given by the large
value of the ratio of the lifetimes $\alpha = \tau_t/\tau_q = 78 \pm 2$
for our
sample \cite{Mironov14}. $\alpha$ is sometimes referred to as the
Dingle ratio.
Here $\tau_t$ is the transport lifetime. It is measured in zero magnetic
field and characterizes the rate of momentum relaxation. The so-called
quantum
lifetime, $\tau_q$, is deduced from the temperature
and field dependence of the Shubnikov de Haas
oscillations. It characterizes the rate at which dephasing occurs. 
A large value of $\tau_t / \tau_q$
is taken to mean that a sample is dominated by the
long-range impurity potential.   While all scattering
processes contribute to $1/\tau_q$, scattering off a long range
potential involves predominantly small angle scattering that has a
limited effect on momentum relaxation. The
corresponding scattering rate, $1/\tau_t$, is smaller than $1/\tau_q$.

The slowly-varying backround potential, set up by the ionized donors,
gives rise to compressible puddles of quasiparticles (QP) and
quasiholes (QH), as indicated in Fig. 2 in the main text. 
The Coulomb interaction, it is assumed, ensures local equilibrium between
excitations at the edges of these puddles and those at the center.
An edge QP (QH) drifts along the equipotential lines surrounding the
puddle, see Fig \ref{fig:Tunnel}. At a saddle point it can cross 
the saddle point via a combination of thermal excitation and
quantum tunneling. The probability of tunneling, as a function of the
closest point of approach of the classical trajectory to the saddle point
$\pm x_0$, was found in \cite{Fertig_Halperin87} 
to be given by
${\cal T}(x_0) = 1/(1+\exp{A})$ where $A=-\pi x_0^2/l_q^2$ with $l_q$
the QP/QH magnetic length. This is the value
predicted by the WKB approximation with $A= \int_{-x_0}^{x_0} Y\cdot
X$, where $Y$ is the momentum in the classically forbidden region.

The conductivity across the saddle is found after computing the
net flow across the saddle and differentiating with respect to the
chemical potential difference:
\begin{equation}
\sigma =  \frac{(qe)^2}{h} \frac{1}{kT}  \int_{0}^{\Delta_s} dE\;  {\cal T} (E-E_s) e^{- E /
kT}.
\end{equation}
Here ${\cal T}$ is written in terms of the energy of the QP
excitation energy, $E$, which defines the point of closest approach, $x_0$,
once a form for the potential is assumed. $E_s$ is the height of the
saddle, and $\Delta_s$ the gap \cite{dABIHMorf11}. We assume that the
potential close to the saddle is given by $V(x,y) = (\Delta_s/2a^2)(-
x^2 +  y^2)$, where $a$ is the typical width of the saddle (see Fig. 2
in the main text). 

Each saddle point acts as an element in a network of resistors for QPs
and as a separate element in a network for QHs \cite{Polyakov94}. The total
conductance of the system is given by $2 (W/L) \sigma$, where
$W/L$ is the effective aspect ratio of the network/sample with $W$ its
width and $L$ its length. The
computation of the resistance network relies on Dykhne's theorem
\cite{Dykhne71}, which states that, for a distribution of resistors in
a network, for which the log of the resistance of the individual
elements is symmetrically distributed about its mean, $\langle \log R
\rangle$, the network resistance
is $\exp \langle \log R \rangle$. We assume that the saddle point
heights, $E_{sp}$ and $E_{sh}=\delta_s-E_{sp}$, are evenly distributed
about $\Delta_s/2$. Once tunnelling is included, $\log R$, is no longer
strictly evenly distributed about its mean (as tunnelling increases
the transition probability for $E_s<\Delta_s/2$ more than for $E_s>\Delta_s/2$).
However, numerical studies of the corresponding resistor networks
shows that the response changes only slightly from that obtained by assuming
the validity of Dykhne's theorem. The deviations from a symmetric
distribution are only significant far from the mean. These elements
act essentially as short or open circuits and do not affect
significantly the network resistance \cite{Asman15}.

\bibliographystyle{apsrev4-1}
\bibliography{SiGeFQHE}